# TOPOLOGY OF THE QUANTUM HALL EFFECT.
# THE MÖBIUS STRIP MODEL.

F. Meseguer

*Instituto de Tecnología Química (CSIC-UPV), Universitat Politècnica de València, Av.*

*Tarongers s/n, 46022 Valencia, Spain*

.

*Here we conjecture on a topological model based on the* Möbius *strip derived from the current distribution at the plateaus of the Quantum Hall Effect (QHE). It can account for the fractional values of the QHE in an easy way.*

Here we will show that edge states for the QHE can be regarded, with new eyes, as a Möbius strip like configuration. It makes Fractional Quantum effect as a natural derivation of the topological distribution of edge states.

The current distribution in a two dimensional electron gas (2DEG) of a Hall bridge at a plateau step goes along the edge of the sample (see Figure 1a).[1] The current distribution in the Hall plateau bar connects the left bottom edge point A to right top edge point B´. If we joint the left edge of the Hall bar to the right one, preserving the current distribution topology we must twist the left edge by $\pi$ before pasting A to A´ and B to B´. It results in a Möbius strip band, like that shown in the Figure 1b.[2] The Möbius strip is the simplest example of a nonorientable manifold. A surface element defined as $d\mathbf{\Omega} = d\mathbf{x} \wedge d\mathbf{y}$ transforms into $-d\mathbf{\Omega} = d\mathbf{x} \wedge d\mathbf{y}$ after a circumnavigation $2\pi$ along the strip as shown in the Figure 1b.[3] In the Möbius strip $d\mathbf{\Omega}$ transforms into itself only when a $4\pi$ circumnavigation is made. $4\pi$ isotropy also appears in the symmetry properties of spinors[4] and it has been related to the Berry phase[5] appearing in graphene[6,7]. Manifolds can be represented by their plane diagram[2] as that shown in Figure 1c for the case of the Möbius strip. A plane diagram is a representation in two dimensions of the manifold, which it is made by cutting the manifold, and then projecting it onto a plane. If we make a transversal cut of the Möbius strip and unfold it, we then obtain the figure 1c. The arrows in both boundaries indicate how we should bind them to reconstruct the manifold. The topology of the QHE currents is better understood in terms of the Möbius strip when we decompose Möbius strip band with the help of oriented 2-simplex.[3] In our case the 2-simplex is an oriented triangle, and the triangulation of the Möbius strip is shown in Figure 1d. Here we can see that triangulation with oriented 2-simplex mimic the Landau orbits and also the current distribution in the Hall plateaus of a 2DEG. So far we have see a topologic model of currents in the QHE. The FQHE has many similarities with the Integer Quantum Hall effect (IQHE) as it also appears in 2DEG samples and shows quantized values of the Hall resistance. However, in this case, Hall plateaus appear at fractional values, for high mobility samples at very low temperatures. Laughling has nicely explained the FQHE,[8] in a many-body interaction framework where Coulomb repulsion play a pivoting role.[1] Later, Halperin showed minima of the cohesive electron energy at odd denominator filling fraction values fractions.[9,1] Here we will show that odd denominator fractions of the FQHE can also be derived by using topological arguments. Let us assume all fractions are developed in a hierarchical manner[1]. Then all odd denominator fractions arise from 1/3 and 2/3. Following this hierarchical rule all even denominator fraction would arise from the 1/2 fraction. Therefore, let us focus on the fractional numbers n=1/2, and n=1/3. In the case of n=1/2 (n=1/3) there is one electron to be shared for each two (three) Landau orbits. For the case of n=1/2, we would have, in the one electron picture, only one half of Landau orbits occupied, the rest being empty. In a many-body description each two Landau orbits would share one electron. From the topology side it would correspond as if the Möbius strip were split along the longitudinal direction into two halves. The amazing result we obtain when doing such operation is not a two-separated Möbius strips half width each one, but a single double-length strip of half-width with two twists instead. The reader can proof such transformation by constructing a Möbius strip with a piece of paper and glue. Then cutting it into two halves along the longitudinal direction. The Figure 2a shows the resulting plane diagram when a longitudinal division of the plane diagram is made. The plane diagram of the Figure 2a does not correspond to a Möbius strip but to a cylinder instead. Therefore the n=1/2 fractional value is not consistent with the Möbius like current distribution topology shown in Figure 1a. From the absence of the topology conditions of the n=1/2, as well as from the hierarchy of the different fractions, where 1/2p, p=1,2 3, etc., all even denominator fractions come from ½ fraction, one can infer they would be absent in the FQHE.

For the case of n=1/3, one third of Landau orbits will be occupied. Here we follow the same procedure as made for the case of n=1/2. After cutting longitudinally the Möbius strip into three parts we obtain two bands, one of them with one twist and the same length as the original one, but one-third thinner. We also obtain a second strip double large and one third thinner but with two twists as shown in the Figure 1b. Therefore the fractional number n=1/3 conserves the current distribution topology. Concerning odd denominator fractions, it is easy to show that filling fraction values as n=1/5, 1/7, 1/9 can be derived following the same procedure as shown for the 1/3 case. By using topology arguments, I have shown that odd denominator fractions like 1/3, 1/5, 1/7, etc are possible fractions candidates for the FQHE. However, even denominator fractions (1/2, 1/4 etc.) are not. Theoretical calculations of Halpering[9] have shown minima for odd denominator filling fractions and maxima for even denominators fractions.

Let us now discuss about the origin of the Möbius strip model for the case of the QHE and the connection to other quantized phenomena like the QSH effect in inverted Quantum Wells as well as that for the case of the graphene. The Möbius strip can be a topological model of pinors.[10] Pinors in (*n-1*) dimensions are equivalent to spinors in *n* dimensions.[11] Electrons (holes) with half integer spin value are governed by Fermi-Dirac statistics, a condition for explaining the FQHE.[1] Therefore the Möbius strip model here used might be connected to the double valuated representation of spinors.[4]

In summary we have shown a topology model of the QHE that may account for the fractions appearing in the FQHE of semiconductor heterojunctions. This model may give hints for understanding experiments of QHE in graphene and also for QSH effect appearing for inverted QW as well as for topological insulators. More experiments would be necessary for corroborating this model.

Acknowledgements.

This work has been partially supported by the following grants  MAT2015-69669-P as well as the regional project from Valencia (Spain), PrometeoII/2014/026

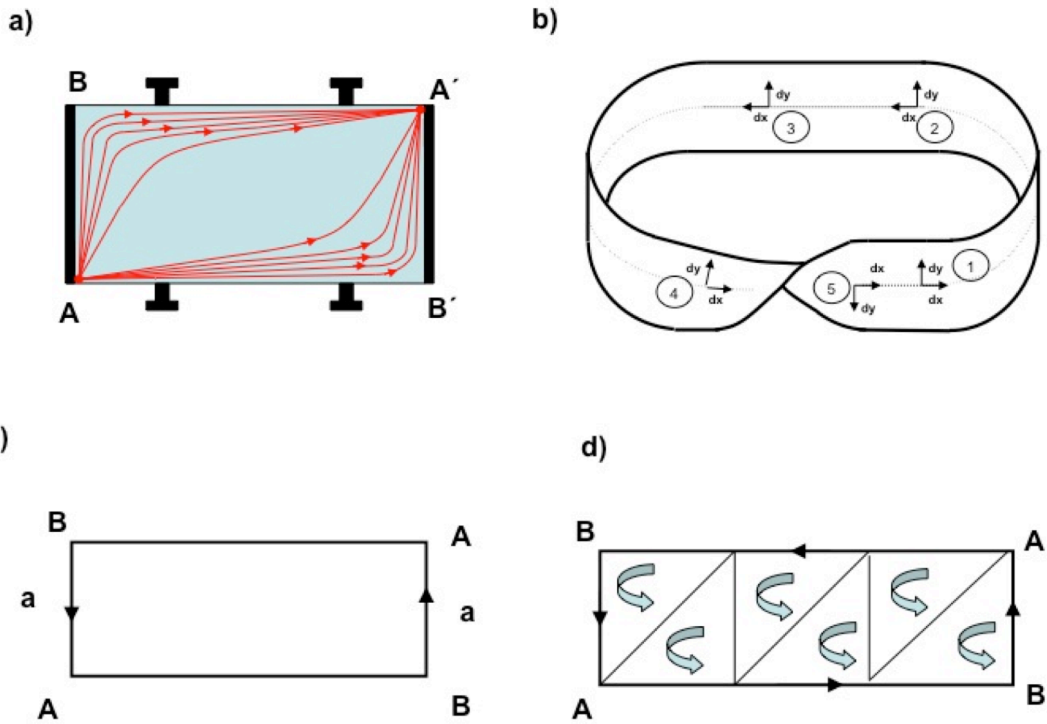

**Figure 1.** Top view of a Hall bar with the current flow distribution at the quantum Hall step (fig. 1a). Möbius strip model showing how a differential surface element $d\Omega = dx \wedge dy$ transforms after a $2\pi$ circumnavigation from step 1 to step 5 (fig. 1b). Plane diagram of the Möbius strip (figure 1c). Oriented triangulation of a plane diagram corresponding to the Möbius strip (figure 1d).

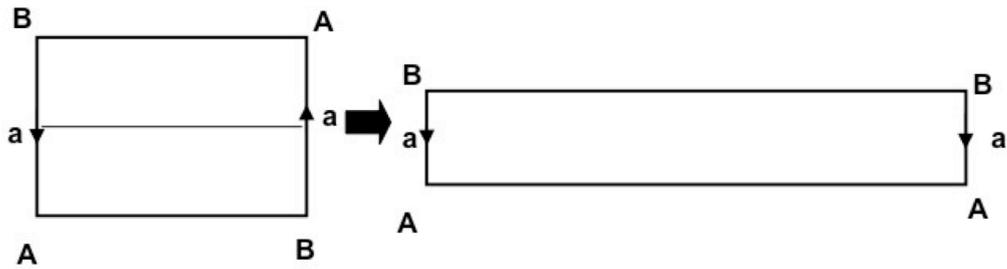

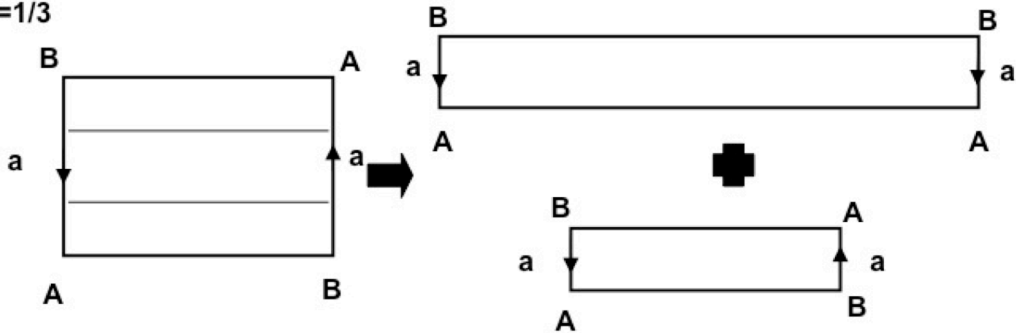

Figure 2

**Figure 2. a)** Plane diagram (left) of the Möbius strip at the filling fraction n=1/2 and the decomposition of the plane diagram (right) into a cylinder plane diagram when the Möbius strip is split longitudinally into two halves (right). **b)** Plane diagram (left) of the Möbius strip at the electron filling fraction n=1/3 and the decomposition of the plane diagram into a cylinder plane diagram 1/3 wide (top right) and a Möbius strip plane diagram (bottom right) one third wide.